\documentclass[12pt,a4paper]{article}
\usepackage{authblk}
\usepackage{tikz-cd}
\usepackage[english]{babel}
\usepackage[latin1]{inputenc}
\usepackage{lmodern}
\usepackage{verbatim}
\usepackage{amsfonts}
\usepackage{amsmath}
\usepackage[T1]{fontenc}
\usepackage{color}
\usepackage[normalem]{ulem}
\usepackage{todonotes}
\usepackage{natbib}
\usepackage{url}
\usepackage[bookmarksnumbered=true, colorlinks=true, citecolor=blue]{hyperref}
\date{}

\title{The Metaphysics of Machian Frame-Dragging}
\author{Antonio Vassallo}
\author{Carl Hoefer}
\affil{LOGOS-BIAP\\University of Barcelona, Department of Logic, History and Philosophy of Science\\ Carrer de Montalegre, 6\\ 08001 Barcelona}

\begin{document}

\maketitle
\begin{center}
\end{center}
\pdfbookmark[1]{Abstract}{abstract}
\begin{abstract}
The paper investigates the kind of dependence relation that best portrays Machian frame-dragging in general relativity. The question is tricky because frame-dragging relates local inertial frames to distant distributions of matter in a time-independent way, thus establishing some sort of non-local link between the two. For this reason, a plain causal interpretation of frame-dragging faces huge challenges. The paper will shed light on the issue by using a generalized structural equation model analysis in terms of manipulationist counterfactuals recently applied in the context of metaphysical enquiry by \citet{471} and \citet{472}. The verdict of the analysis will be that frame-dragging is best understood in terms of a novel type of dependence relation that is half-way between causation and grounding.
\end{abstract}
 
\section{Introduction}
It is virtually impossible to address the problem of the origin of inertia in spacetime theories without mentioning Ernst Mach and his ``The Science of Mechanics'' \citep{110}. Indeed, Mach's views on inertia have been discussed and analyzed at length in the philosophical literature. The consensus among Mach's commenters is that his discomfort with the origin of inertia in Newtonian mechanics comes from epistemological considerations filtered through empiricist inclinations. For Mach, it is highly unsatisfying to link inertia to Newtonian absolute space for the simple reason that such a space is unobservable. If the aim of physics -- and science in general -- is to provide a picture of the world based on experience, then there is no place in this picture for elements that elude experience in one way or another. 

At this point, however, the consensus breaks down as how much further Mach pushed these considerations. On a prudent reading, it seems that Mach simply argues for a mere redescription of the role of inertia in classical mechanics, that is, that inertia should not be referred to absolute space, but to a suitably defined reference frame (e.g., in the case of Newton's bucket experiment, the inertial effects arising should be referred to the distant background of fixed stars). A more radical reading instead suggests that Mach has in mind not just a mere redescription of classical mechanics, but a brand new physical theory in which inertia originates from the overall distribution of masses in the universe \citep[see][section 8]{483}. It is no mystery that Einstein saw this ``radical Mach'' as one of his main inspirations in his quest for general relativity (GR). Although there are still disputes about whether GR fully complies with a radical interpretation of Mach's views \citep[see for example][chapter 3]{493}, it is a well-established fact that, under certain physical conditions, the theory predicts that local inertial frames are determined to a certain degree by the surrounding material distribution. This is the case for frame-dragging effects, such as the Einstein-Lense-Thirring effect, according to which a gyroscope orbiting a huge rotating mass distribution will show a precession that is directly related to the angular momentum of the distribution \citep{494}. 

In a nutshell, frame-dragging effects imply that the axes of a local inertial frame are not fixed and independent of the surrounding material distribution, which in turn means there is no such thing as an absolute space acting as a fixed reference for an inertial compass. Section \ref{frdr} will provide a very brief introduction to rotational frame-dragging effects in GR, from modest Machian effects implying that local inertial frames are determined both by the state of motion of surrounding matter distribution and the condition of asymptotic flatness at infinity, to fully Machian effects dictating that inertial frames are defined with respect to the overall matter distribution in the universe. The key point to be highlighted is that, in any case, the determination relation established does not depend on time. This apparently means that local inertial frames would instantaneously ``feel'' any change in the surroundings' state of motion, irrespective of how far away in space such change is triggered. This may be a source of unease for those inclined to consider frame-dragging as a result of some sort of physical interaction. In fact, the absence of a retarded mechanism underlying these effects might lead them to consider frame-dragging as an instance of action-at-a-distance. This would in turn raise delicate questions for metaphysicians who want to read a causal dependence off from the frame-dragging mechanism, for they would then apparently be forced to accept some kind of superluminal causation. Although this would not by itself be a fatal blow to a causal interpretation of frame-dragging (after all, one might simply take it to show that in GR there is superluminal causation), still one might take this controversial issue as a motivation to pursue an alternative metaphysical analysis. This is exactly what is done in section \ref{sem}, where the framework of structural equation models (SEM) will be discussed. The framework helps to analyze the dependency relations underlying a set of correlations by (i) constructing a mathematical model of such dependencies and (ii) counterfactually testing the model in a manipulationist fashion. This tool has been so far used mainly in social science to extract causal information from huge datasets but, recently, some metaphysicians have noted that the framework works in principle with any kind of determination relation, and in particular with grounding. The aim of this paper is to apply the SEM framework to shed light on the issue of finding the ``right'' dependence relation depicted by frame-dragging: the results of our analysis are presented in section \ref{frcs} and further discussed in section \ref{five}.

\section{Rotational frame-dragging effects in GR}\label{frdr}
The conceptual path that leads to general relativistic frame-dragging can be traced back to Newton's bucket argument. This thought experiment was meant to show that the relative rotation of water with respect to the bucket's walls was neither necessary nor sufficient to ground the centrifugal forces responsible for the concavity of the water's surface. Newton's conclusion was that centrifugal forces arose as a result of water rotating with respect to absolute space, since all possible inertial frames were tied to this latter entity. Mach famously replied to this challenge by pointing out that all we can \emph{observe} is that the curving of the water's surface is not determined by its state of motion with respect to the walls of the bucket, but is determined by its state of motion relative to the frame of the fixed stars. We do not know, Mach went on to say, whether the curving of the water's surface would be just the same if all the matter of the fixed stars were removed from the universe, nor whether a curving effect might in fact be produced by mere relative rotation with respect to the walls of a bucket, if the latter were ``several leagues thick''. GR vindicates Mach's point of view on dragging effects in that (a) in asymptotically flat universes any local inertial compass would nonetheless feel the presence of a surrounding rotating matter distribution, and (b) in non-asymptotically flat universes inertial frames appear to be, in some models at least, completely determined by the overall mass distribution. With these facts in mind, we can say that dragging effects in GR come in different degrees of Machianity. In the following, we will review the most important results, placing them in an ascending scale of Machianity from (a) to (b), without any pretense to be exhaustive or mathematically rigorous.

The less Machian frame-dragging case was first discussed in the early years of GR, by Einstein, Lense, and Thirring (here we draw from \citealp{495}, section 6.1, and \citealp{27}, sections 18.1 and 19.1). Consider a slowly rotating ideally spherical body with mass $M$ and angular momentum $\mathbf{J}$ in a stationary, asymptotically flat spacetime. In the weak field limit we can use linearized gravity, that is, we can decompose the metric $g_{\mu\nu}$ into a flat Minkowski background $\eta_{\mu\nu}$ plus a perturbation term $h_{\mu\nu}$ (throughout the paper we will set $G=c=1$):
\begin{equation}\label{lin}
g_{\mu\nu}\approx\eta_{\mu\nu}+h_{\mu\nu},
\end{equation}
and define:
\begin{equation}\label{lin2}
\bar{h}_{\mu\nu}\equiv h_{\mu\nu}-\frac{1}{2}\eta_{\mu\nu}\left(\eta^{\alpha\beta}h_{\alpha\beta}\right).
\end{equation}
The general solution of the linearized field equations will then be:
\begin{equation}\label{lin2}
\bar{h}_{\mu\nu}=4\int\frac{T_{\mu\nu}(\tau,\mathbf{x'})}{|\mathbf{x}-\mathbf{x'}|}d^3x',
\end{equation}
where $\tau=t-|\mathbf{x}-\mathbf{x'}|$ is the ``retarded time'', which models the fact that big enough perturbations of the background would propagate from the source at a finite velocity as gravitational waves. However, in the present case, velocities are too low to produce such big perturbations. Moreover, in general, by expanding $h_{\mu\nu}$ in powers of $\frac{\mathbf{x'}}{|\mathbf{x}-\mathbf{x'}|}\equiv\frac{\mathbf{x'}}{r}$, we see that the higher-order radiation terms die out as $\frac{1}{r}$ so, if the radius of the body is big enough, no retardation effect will be seen.

In this context, the infinitesimal $0i$ $(i=1,2,3)$  components of Einstein's field equations can be written in the Lorenz gauge as:
\begin{equation}\label{lensth0}
\delta \bar{h}_{0i}=16\pi\rho v^i,
\end{equation}
with $\rho$ the density and $v^i$ the linear $3$-velocity of the mass distribution. The solution will be:
\begin{equation}\label{lensth1}
\bar{h}_{0i}=-4\int\frac{\rho(\mathbf{x}')v^i(\mathbf{x}')}{|\mathbf{x}-\mathbf{x}'|}d^3x.
\end{equation}
In the end, by passing to spherical coordinates, the metric outside the body will approximately be (in the appropriate (Kerr) gauge, where $\mathbf{J}=(0,0,J)$):
\begin{equation}\label{lensth2}
\begin{split}
ds^2=-\left(1-\frac{2M}{r}\right)dt^2+\left(1-\frac{2M}{r}\right)^{-1}dr^2+r^2\left(d\theta^2+sin^2\theta d\phi^2\right)-\\-\frac{4J}{r}sin^2\theta dt d\phi.
\end{split}
\end{equation}
The off-diagonal term $\bar{h}_{t\phi}$ in \eqref{lensth2} comes from \eqref{lensth1} and from the fact that $\mathbf{J}=\int\mathbf{x}\times(\rho\mathbf{v})d^3x$, and can be considered some sort of potential in analogy to the magnetic potential in electrodynamics. More precisely, it is a ``dragging potential'' because it can be shown that a gyroscope orbiting the massive body would precess -- with respect to an observer at infinity -- with angular velocity:
\begin{equation}\label{lensth3}
\boldsymbol{\Omega}=-\frac{1}{2}\boldsymbol{\nabla}\times\mathbf{A},
\end{equation}
where $\mathbf{A}=\left(0,0,-\frac{2J}{r}sin^2\theta\right)$ and $\boldsymbol{\nabla}$ is the covariant spatial derivative. \eqref{lensth3} is thus a gravitational analogue of Faraday's law of induction in electrodynamics. Without entering into further mathematical detail, we already see what is Machian about the result \eqref{lensth3}: since the gyroscope determines the axes of a local inertial frame, the state of motion of the mass distribution influences such a determination in the vicinity of its surface. Note that $\boldsymbol{\Omega}$ varies considerably at different locations. For example, near the poles the inertial frames rotate in the same direction as the massive body, while near the equator the inertial frames rotate in the opposite direction. We also see what is \emph{not} Machian here (basically, everything else): the rotation of the massive body and the precession of the gyroscope are referenced to the \emph{infinity} fixed ab initio without reference to any mass distribution. Formally, this means that \eqref{lensth3} is solved modulo a constant of integration that the asymptotic flatness condition fixes to be $\boldsymbol{\Omega}(\infty)=0$.  In this sense, for $r\rightarrow\infty$ we get the ``true'' inertial frame, which suspiciously looks like an absolute space in the Newtonian sense.

A slightly more Machian dragging effect arises if we consider a slowly rotating massive spherical shell. To determine the metric inside this body, we just note that, in the case $\mathbf{J}=0$, this metric would be the flat Minkowskian one $\eta_{\mu\nu}$. Hence, for $\mathbf{J}\neq0$, in the weak field limit we can apply again \eqref{lin}, $h_{\mu\nu}$ again being a rotational perturbation small enough that the resulting metric is stationary (in order to avoid higher order contributions form gravitational waves). With this machinery in place, we can apply the same reasoning above, thus finding that the axes of the gyroscope near the center of the shell would be dragged with respect to infinity. This situation is slightly more Machian than the previous one because, in the reference frame of the gyroscope, the metric inside the shell appears flat, thus hinting at the fact that the matter distribution defines ``its own'' inertial frames inside the shell. However, the Machianity of the situation ends here, because the rotations involved are not defined relationally but refer, again, to infinity. Furthermore, the approximations used in the model make it too much of a toy model. For example, we need to impose physically implausible conditions on the stress-energy tensor in order to keep the radius of the shell constant, this in turn makes it troublesome to ``connect'' the metric inside the shell with that on the outside.

More physically realistic models of slowly rotating mass shells with a time dependent radius (in order to model their expansion/contraction) are due to \citet{496,497,498}. Roughly, these authors apply a slightly more sophisticated perturbation approach to the outside of the shell. Given that, by Birkhoff's theorem, the metric outside a static spherical distribution of matter in an otherwise empty universe is the Schwarzschild one, they start by a rotationally perturbed version of it, which can be generically written as:
\begin{equation}\label{lensth4}
ds^2=e^{2\alpha}dt^2-e^{2\beta}dr^{2}-r^2\left[d\theta^2+sin^2\theta\left(d\phi-\omega dt\right)^2\right],
\end{equation}
where $\omega=2\frac{J}{r^3}$, and $e^{2\alpha}$ and $e^{2\beta}$ are appropriate functions of the shell's mass determined by the field equations plus the state equation for matter. The metric inside the shell is taken to be flat, although written in rotating coordinates depending also on the radius of the shell. In this way, the authors are able to give a full description of the system, including the connecting region on the shell. The angular momentum $\mathbf{J}=(0,0,J)$ within a region $r$ depends on the $t\phi$ component of the stress-energy tensor through:
\begin{equation}\label{lensth4a}
J=\int_{0}^{r}\int\int T_{\phi}^{t}\sqrt{-g}d\theta d\phi dr.
\end{equation}
Again, we do not need to go too much into technicalities here: the interesting point for us is that all these authors recover frame-dragging results that are a generalized form of \eqref{lensth3}:
\begin{equation}\label{lensth5}
e^{-\alpha}\boldsymbol{\Omega}=-\frac{1}{2}e^{\alpha}\boldsymbol{\nabla}\times\mathbf{A},
\end{equation}
where the factor $e^{-\alpha}$ accounts for the increase of rotation rate of the shell due to the gravitational slowing of the clocks close to the shell with respect to those at infinity.

\citet{499} further generalize these results by considering $N$ nested freely falling shells, the metric between any two shells being of the form \eqref{lensth4}, and taking the continuous limit (i.e. $N\rightarrow\infty$). They find the interior variation of the metric's perturbation $\omega$ to be:
\begin{equation}\label{lensth6}
\frac{\partial\omega}{\partial r}=-6e^{\alpha+\beta}\frac{J}{r^4}.
\end{equation}
If we assume asymptotic infinity, so that $\omega\rightarrow0$ for $r\rightarrow\infty$, we can integrate \eqref{lensth6} to get the overall perturbation distribution over space:
\begin{equation}\label{lensth7}
\omega=2\left(\frac{W}{r^3}J+\int_{r}^{\infty}\frac{W}{r'^3}\frac{\partial J}{\partial r'}dr'\right),
\end{equation}
where $W$ is an appropriate weight function that depends on $e^{\alpha+\beta}$ and $r$. Roughly, \eqref{lensth7} is a measure of how much dragging comes from any spatial region of the model. Hence, this model nicely describes a case where local inertial frames are partially determined by the overall matter distribution in the universe. This is the most Machian scenario still falling in the category (a).

The subsequent step is to render the model even more Machian by assuming that there is no infinity region, i.e. by imposing that the universe is closed. In this case, it is more useful to define a new coordinate $0<\chi<\pi$ such that $sin\chi=\frac{r}{r_{max}}$, $r_{max}$ being the maximum size of the universe. Simple geometrical considerations lead to a new version of \eqref{lensth6} appropriate for the closed case:
\begin{equation}\label{lensth8}
\frac{\partial\omega}{\partial \chi}=-6e^{\alpha+\beta}\frac{\partial r}{\partial\chi}\frac{J}{r^4}.
\end{equation}
When integrating \eqref{lensth8} in order to get the closed universe equivalent of \eqref{lensth7}, we have to keep in mind that there is no more boundary condition $\omega(\infty)=0$ because there is no infinity. The only thing we can do is to integrate with respect to an arbitrarily chosen point $\chi_*$:
\begin{equation}\label{lensth9}
\omega-\omega_*=2\left(\frac{W_*}{r^3}J+\int_{\chi}^{\chi_*}\frac{W_*}{r'^3}\frac{\partial J}{\partial \chi'}d\chi'\right).
\end{equation}
Equation \eqref{lensth9} tells us that, since there is no privileged inertial frame at infinity, the angular momentum distribution just contains information about the \emph{relative} rotations of inertial frames at different points. The icing on the cake is given by a mathematical result proven by the authors in this context, which states that the total angular momentum of a closed universe is necessarily zero. This is the most Machian setting among those considered so far, in which dragging effects arise. Equation \eqref{lensth9} comes very close to Mach's idea encompassed in his reply to Newton, i.e. that the concavity on the surface of a rotating mass of water was just due to the dragging of the water's inertial frame by the relative counter-rotation of the surrounding matter in the entire universe. 

These results can be further generalized by considering small perturbations of a FLRW metric:
\begin{equation}\label{FLRW}
ds^2=dt^2-a^2(t)\left[\frac{1}{1-kr^2}+r^2\left(d\theta^2+sin^2\theta d\phi^2\right)\right],
\end{equation}
$a(t)$ being a scale factor and k=-1,0,1 representing the (constant) spatial curvature. The perturbed metric $\tilde{ds}^2$ will feature off-diagonal elements corresponding to such perturbations. The key point to consider is that, when we want to ``project'' $\tilde{ds}^2$ on the unperturbed background $ds^2$, we have a number of gauge degrees of freedom associated with the background's underlying symmetries. In this particular case, the unperturbed metric exhibits spatial homogeneity and rotational symmetry which, in particular, means that the associated Killing vectors are $3$-dimensional. In this context, Noether's theorem implies that any conserved quantity at a given instant of cosmic time $t$ features -- at the first perturbational order -- only the $0i$ $(i=1,2,3)$ components of the stress-energy tensor and, hence, only the ``spatial constraints'' of the Einstein's field equations are involved. From all of this, we can calculate the off-diagonal $h_{0i}$ terms in $\tilde{ds}^2$ as functions of $3$-Killing displacements $\xi_{ai}$ $(a=1,\dots,6)$ by integrating the set of equations:
\begin{equation}\label{disp}
\delta h_{0i}=f^a\xi_{ai}.
\end{equation}
\eqref{disp} is a general form that includes \eqref{lensth7} and \eqref{lensth9} as particular cases, $f^a$ being a six-component integration constant. Note that no timelike derivative appears in this relation. In the case of an open universe, $f^a$ is fixed by the boundary conditions at infinity (as in the case $\omega(\infty)=0$ in \eqref{lensth7}) but, if the universe is closed, some freedom will remain as to choose $f^a$, meaning that only relative motions are definable, in the same Machian spirit as \eqref{lensth9}. This latter result has been later extended -- albeit in very particular gauges -- to \emph{any} FLRW cosmology, thus entirely fulfilling Mach's ideas \citep[see][]{500}.

To sum up, rotational frame-dragging effects are perhaps best understood if we bring in again the analogy with electrodynamics, which is apparent in  \eqref{lensth3}. Such effects are induced by a dragging potential encoded in the off-diagonal terms representing small perturbations of a background metric (be it Minkowski, Schwarzschild, or FLRW). The very general form of these perturbations is \eqref{disp}, whose right-hand side is determined by the mass-energy distribution in the region of interest (as, for example, in \eqref{lensth0}). In all these cases, however, the magnitude of $\delta h_{0i}$ is so small that no retarded action of the form \eqref{lin2} will arise because no substantial gravitational radiation that literally carries the perturbation will be produced. This is an obvious disanalogy with the electrodynamic case, where physically realistic potentials -- especially in long-range interactions -- are retarded. The immediate consequence of this lack of retardation in the general relativistic context is that introducing any $\delta$-sized change in the mass-energy distribution of a system will \emph{instantaneously} alter the magnitude of the dragging effects felt by local inertial frames. For example, if we take a spherical shell of matter centered on a point $P$ in a FLRW universe and give it a small rotation about $P$, the result of the perturbation is ``felt'' instantaneously at $P$.  As we noted above, this ``effect'' is instantaneous on a constant-$t$ hypersurface in the model due to the fact that its existence is derived from the spatial constraint equations, which can be thought of as one part of Einstein's field equations (note, \emph{en passant}, that no violation of general covariance is implied, i.e. frame-dragging effects do not identify a \emph{privileged} foliation of spacetime).

This is of course a bit of a problem -- especially in a cosmological context -- if we wish to say that local inertial frames are \emph{causally} affected by the surrounding matter distribution. In fact, causation is usually understood as a diachronic relation (causes precede their effects) but, even if we are willing to relax this condition enough to admit at least synchronic causation, we are left with something that looks a bit like spooky action-at-a-distance. And even if we bite this bullet, still we will face the challenge of explaining whether, and if so why, this action amounts to -- or does not amount to -- some kind of superluminal physical influence.

The problems in this context are made even worse by the fact that no well-established philosophical account of causation is able to provide a neat analysis of dragging dependencies. Conserved quantity approaches \`a la Salmon-Dowe are notoriously ill-defined in GR, and also causal property theories such as Alexander Bird's dispositional monism are at odds with many foundational aspects of the theory \citep[see][and references therein]{126}. The situation is not better for counterfactual theories such as Lewis' or  Woodward's, given that the GR's dynamical geometry gives no well-established standard to evaluate a counterfactual change against the actual situation (see \citealp{487}, for a general discussion, and \citealp{504}, for the specific case of frame-dragging). The first difficulty regards a counterfactual situation involving a local change that leaves everything else untouched. In this case, even if we are working with a model of GR that admits a well-posed initial value formulation and we express this ``change'' as a tiny modification of the metrical and material properties in a small neighborhood on the initial spacelike surface, such a change would be enough to violate the field equations, which constrain the happenings on each surface, thus making any attempt to evolve the situation to see ``what would had happened if we had made this small change, leaving everything else as is in the actual world'' far-fetched. The second problem regards testing counterfactuals by comparing two models of the theory, corresponding to the actual and counterfactual situations. In this case, the lack of a standard implies that the way we choose the counterfactual model is always somewhat arbitrary. This point becomes evident if we consider the counterfactual situation where all the masses in the actual world would vanish (decay into the true Higgs vacuum is a concrete possibility...). We might naturally be inclined to say that, if that happened, the Minkowski solution would be the best candidate to model the situation. However, Minkowski spacetime is not the only vacuum solution of Einstein's field equations, and no compelling mathematical argument can be made that singles out this particular solution over the others. Moreover, as is pointed out in \citealp{499}, if one were to approach modelling this counterfactual situation via a series of FLRW models with ever-smaller cosmic mass, then if one starts from a closed/finite universe, as the total cosmic mass gets smaller and smaller the spacetime becomes smaller, shorter-lived, and more highly curved, eventually (in effect) \emph{vanishing} in the limit as cosmic mass goes to zero, rather than turning into Minkowski spacetime (or any other infinite empty spacetime of GR).

The ambiguity problem affects our treatment, for example, in the Einstein-Lense-Thirring case. As a matter of fact, one can consider the external field of a spinning nearly spherical massive body in an otherwise empty universe (Kerr model) and then test the counterfactual ``If the body had not spun, the local inertial frame $K$ at a distance $|r|$ from the body's surface would not have precessed'' against a Schwarzschild model for a body with same mass and radius but no angular momentum, thus finding that the counterfactual is true. In this case, our choice would be more robust because of Birkhoff's theorem. However, the problem is still there, and manifests itself in the fact that there is no objective well-established trans-model identity criterion that helps us point at $K$ in the starting model and say that the orientation of the $z$-axis of the very same $K$  in the second model does not precess. Usually, physicists make sense of such counterfactuals by stipulation: they map the perturbed space (the Kerr model, in this example) on the unperturbed background (here, the Schwarzschild model) in a way that fixes for any point $P$ in the background its ``perturbed'' counterpart $P'$. This fixing procedure basically amounts to the condition that $P$ and $P'$ are always flagged with the same coordinate value. In this way, any choice of coordinates on the background will immediately fix the coordinates in the perturbed space. Thanks to this stipulation, we immediately see that we can evaluate the above counterfactual without particular worries. However, this approach to trans-model identity just works on a case-by-case basis and, therefore, cannot be adopted as a formal backbone of a general counterfactual analysis of causation.

In the following section we will consider a different approach to the analysis of causation, one which, in our opinion, is a promising framework for investigating the nature of dependence relations in GR.

\section{Structural equation model analysis}\label{sem}
The umbrella title ``structural equation modelling'' designates a set of mathematical tools developed as early as the 1960s in parallel with the rise of database management systems. The main scope of these tools is to analyze distributions of data by best-fitting them in graph structures that explain some  ``mechanism'' (in a loose sense) of interest underlying these distributions. So far, these models have been mainly used in sociology and psychology to, e.g., test the hypothesis that intelligence (under a certain operational measure) strongly influences academic performance. It is worth noting that this framework is tentative in nature and, as such, it progresses on a trial-and-error basis. First, a certain graph structure is proposed, which describes how data - sorted out by different kinds of variables - are related based on some explanatory hypotheses, and then it is tested, e.g. by statistical methods, how well actual data fit the network of dependencies posited: if the results are unsatisfactory, a modification of the starting hypotheses is made and the entire process is performed again.

The SEM framework naturally lends itself to causal analysis and in particular to the evaluation of causal inferences \citep[See][for one of the first attempts to spell out this approach in detail]{501}. For our purposes, the following example will suffice \citep[see][section 2, for an extensive presentation and discussion of the framework in a philosophical context]{471}. Imagine we have a dataset that shows a systematic correlation between the members of two distinct samples $X$ and $Y$, and we hypothesize that there is some sort of causal mechanism that makes it the case that $Y$ depends on $X$. The starting point is to define an ``endogenous'' variable $y\in Y$ representing the dependent condition and an ``exogenous'' variable $x\in X$ representing the independent condition. We further require that these variables take values from an appropriate set, for example the binary set $\left\{0,1\right\}$, ``$0$'' meaning that the condition does not obtain and ``$1$'' meaning that it obtains. The next step is to provide a set of formal relations (e.g. equations) showing how $y$'s value has to be evaluated on the basis of $x$'s value. We can symbolize this relations by $y\overset{\leftarrow}{=}f(x)$, where the symbol ``$\overset{\leftarrow}{=}$'' makes the conjectured direction of dependence explicit. Hence, given an assignment of value $a$ to $x$, the model gives the corresponding value $f(a)$ for $y$. The intuitive interpretation of this formal dependence is straightforward: the ``wiggling'' of $x$ always (if $f$ is deterministic) triggers a corresponding change in $y$ in a ``$f$'' way, but the opposite does not hold. We immediately see that the best way to render this analysis of dependence in terms of wiggling is by using interventionist counterfactuals: ``If an intervention on $x$ had set its value to $a$, then $y$ would have taken on the value $f(a)$'' \citep[see][for a survey of the interventionist framework]{502}. This also makes it clear why the independent variable is called exogenous: it is the one that is subjected to any process external to the system, i.e. an intervention. Finally, the model is tested against the dataset to see how accurately $f$ captures the correlations pattern between the two samples. Although it most naturally and powerfully applies to generic or ``type'' causation, the method can be used to model both generic and singular (``token'') cases of causation. In this way, the SEM framework permits both qualitative and quantitative causal analysis, and can be easily implemented into causal search algorithms (as in the case of Ramsey and Winberly's Tedrad project, whose codebase is freely available on GitHub). It is instructive to consider how the framework analyzes a textbook case of direct causation, that is, the shattering of a window caused by the throwing of a stone.

In this case we have two variables: ``$T$'' formalizing whether or not the fact that a (sufficiently heavy) stone is thrown obtains, and ``$S$'' formalizing whether the fact that the window shatters obtains. Both variables take values from the binary set $\left\{0,1\right\}$. Based on everyday experience, we hypothesize that $S$ causally depends on $T$ and we model this dependence as $S\overset{\leftarrow}{=}T$. The graph corresponding to this model is:
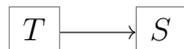
\begin{center}
\begin{tikzcd}[cells={nodes={draw=gray}}]
T \arrow[r, black]
& S
\end{tikzcd}
\end{center}

The model does not need any further tweaking or supplementation to avoid an incorrect back-tracking conclusion ($S \rightarrow T$), as the Lewisian analysis of causation does, because it has the ``right'' counterfactual patterns already built in. As such, this framework is more parsimonious than the Lewisian one, because it dispenses with the talk of possible worlds, small miracles, and the like. It might be objected that this framework, contrary to the Lewisian spirit, is not reductive since the notion of intervention, which plays a key role in the evaluation of the causal link, presupposes an underlying causal process of the same nature as the one analyzed. While we agree that the SEM framework is non-reductive with respect to causation, we do not see it as a problem with the consistency  or the reliability of the inferences drawn in this context since any intervention on $T$ does not necessarily presuppose that we already have causal information about the relationship we want to characterize. Also, here we can see explicitly that the model accounts equally well for the case where \emph{a} stone is thrown through \emph{a} window (type) and for the case where Leonardo throws a stone through Monika's window (token).

Now imagine that there are two stone throwers instead of one. Clearly, the above model would become inaccurate and should be supplemented with a second exogenous variable $T'$ and a second dependence equation $S\overset{\leftarrow}{=}T'$. The correct graph would thus be:
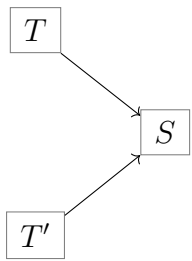
\begin{center}
\begin{tikzcd}[cells={nodes={draw=gray}}]
T \arrow[dr, black] & \\
                              & S \\
T' \arrow[ur, black] &                            
\end{tikzcd}
\end{center}
 
This shows how the framework easily handles cases of causal overdetermination. Similarly, the framework can account for cases of causation by absence: roughly, the dependence of $y$ on the absence of $x$ is modelled by the structural equation $y\overset{\leftarrow}{=}1-x$.
 
It is important to stress the fact that there is no upper limit to the complexity of a model in this context. The graph can have (infinitely) many links and also a branching structure. Graphs can even be cyclic, to model situations (not uncommon in social sciences) where variables can mutually influence one another, or feedback loops exist.  It is common, however, to only consider directed \emph{acyclic} graphs (DAGs) meeting two consistency constraints: (i) that an intervention on a certain variable influences all the other variables downstream but no variable upstream, and (ii) that the topology of the graph is not closed. This restriction is particularly natural when modelling token causation.

The SEM framework has recently attracted the attention of the metaphysical community because it can be easily generalized to any kind of dependence relation that is generative or directed in nature and can be modelled by a partial ordering, grounding being exactly one of these relations. A. Wilson \citep{472}, shows that the analysis of all the major cases of causation can be replicated for grounding. For example, the stone/window case can be translated using the same variables and the same structural equation to the case of  the existence of Socrates grounding the existence of singleton Socrates or, in general, the existence of singletons being grounded in the existence of their respective members. According to Wilson, the only difference between the stone/window case and the member/singleton one lies in the mediating principles involved, that is, the set of background conditions invoked to justify the model. In the first case but not in the second a subset of laws of nature is invoked (note how this law-based demarcation criterion does not mention or imply that causation has to be a diachronic relation). For this reason Wilson renames causation as ``nomological causation'' and grounding as ``metaphysical causation'', going further in arguing that, in fact, standard causation and grounding are at most two species of a genus relation ``causation'' taken as primitive. However, in order for Wilson's unifying framework to work, two slight modifications of the original SEM scheme are in order. First of all, we need to liberalize the notion of intervention. Grounding counterfactuals involve ``metaphysical interventions'', some of which are impossible. Moreover, given that we accept metaphysically impossible interventions, we need to adopt a non-standard semantics of counterpossibles which does not treat them as trivially true. For example, for the structural model $S\overset{\leftarrow}{=}T$ ($T$ being whether Socrates exists, and $S$ being whether singleton Socrates exists) to work, we have to evaluate the counterpossible ``if an intervention had prevented singleton Socrates from existing, then Socrates would not have existed'' as false \citep[see][]{503}.

For our purposes we do not need to go as far as accepting Wilson's (controversial) unification thesis, but we will take on board the view that the SEM framework, including Wilson's suggestion to look for the presence of lawlike background conditions, is a powerful tool to analyze the nature of dependence relations, especially where standard analyses are in trouble, such as in the case of frame-dragging in GR.

\section{Frame-dragging as a non-standard determination relation}\label{frcs}
We ended section \ref{frdr} by pointing out that counterfactual theories of causation are in trouble in the GR context. One can thus ask what progress we really made by adopting SEM's framework, given that it heavily relies on counterfactuals. In reply we note that, first of all, in this context the truth values of counterfactuals are not evaluated by comparing possible worlds, but by computing the values of the structural equations. In section \ref{frdr} we said that in GR it is not so straightforward to consistently and non-arbitrarily single out a model that represents a counterfactual situation to a given one (recall the disappearing masses case). This puts possible worlds semantics in trouble because it undermines the notion of ``nearness'' of possible worlds (which vacuum possible world is closest to ours?). In our case we do not need any metaphysical notion of similarity among possible worlds to evaluate counterfactuals, we just need the posited structural relations among the relevant variables of the model. This shifts the arbitrariness involved in the evaluation of counterfactuals from a metaphysical to a methodological perspective, which is in fact closer to the way physicists engage in counterfactual reasoning.

The problems with interventionist counterfactuals are also mitigated because our counterfactual semantics is much more liberal than the standard interventionist one, allowing for any kind of physical and metaphysical intervention, possible or impossible. The worry that, say, a physically impossible intervention would render the analysis void because it would violate Einstein's field equations is unfounded, since the variables we intervene upon are part of the structural equations, whose adoption is of course justified by invoking the field equations, but which are nonetheless distinct from these field equations. In other words, by setting physically or even metaphysically impossible interventions on the variables, we are not feeding garbage into the theory's dynamics, thus getting garbage in return, but we are just testing the set of relations internal to our structural model. It might be the case that, in the end, the model turns out to be at odds with GR in one way or the other (e.g. local curvature dropping to zero when mass is increased), which only means we have to modify the structural model.

To make this point more vivid, let us consider first the problem of trans-world identification. Clearly, the use of this methodology would dramatically ease the evaluation of counterfactual changes. Indeed, now it is no more matter of looking how, say, the \emph{very same} region $S$ of spacetime in a possible world $W$ looks in a near world $W'$,  because any counterfactual change of $S$ translates into the correspondent change of value of the variable $s$ representing $S$ in the corresponding structural equations. Further, let us consider counterfactuals involving small local changes in a model of GR that leave the rest of the universe untouched. In this case, all we have to do is to take an appropriate (e.g. fine-grained enough) structural model, ``zoom in'' the particular situation by looking at the variables that describe the local state of affairs, change their values as required, and then see if and how this change affects the ``global'' variables in the structural equations. Obviously, we cannot perform such a manipulation directly on a solution of Einstein's field equations. However, we can use the structural model to ``transition'' from the starting solution of the field equations (compatible with the starting values of the variables figuring in the structural equations) to a new one which encloses the changed state of affairs. The necessary condition is that the structural model be fully compatible with the laws of GR, otherwise the changed state of affairs fixed by the new values of the structural variables might not be encoded in any solution of the field equations. Note that this ``transitioning'' between solutions of GR is the SEM counterpart to Lewisian possible worlds' vicinity. Now, however, the similarity comparison between different states of affairs is not justified by metaphysical considerations, but just by pragmatic ones (i.e. that the structural model linking the two solutions works well in capturing the ``backbone'' of dependencies for both situations). It is easy to see how this mitigates also the issue of evaluating the counterfactual situation in which all the masses in the actual universe disappear: just construct an informative enough structural model that rightly captures the dependencies in the actual world, set the material variables in the dependence chain to zero, and see which vacuum model of GR is compatible with this new situation.


The above discussion is very schematic and is meant to hint at the fact that the advocated SEM framework fares better than standard frameworks, at least in cases where it is already clear what one is looking for (the frame-dragging case being one of those, as we will see in a moment). However, we are not claiming that the generalized SEM framework is the panacea for the troubles with counterfactual analyses of dependencies in GR. This line of research is still in its infancy, so a detailed assessment of how well the SEM framework fares \emph{in general} in GR is still to come.

Having argued that the SEM framework is a promising approach to counterfactual reasoning in GR, we can now focus on the concrete case of rotational frame-dragging, and see if Wilson's analysis in terms of metaphysical/nomologic dependence helps us in judging whether causation is involved. For sure, by using Wilson's law-based characterization of dependence, the question of whether the dependence relation between the angular momentum $\mathbf{J}$ of the rotating mass and the angular velocity $\boldsymbol{\Omega}$ of the precessing local frame is instantaneous or retarded becomes irrelevant to the causal verdict: $\boldsymbol{\Omega}$ would causally depend on $\mathbf{J}$ just in case we have to invoke the laws of nature in order to justify the structural equation relating them. Put in these terms, the verdict seems to be trivial: yes, $\boldsymbol{\Omega}$ and $\mathbf{J}$ are causally related because in the analysis of the dependence relation involved we must invoke Einstein's field equations. However, this verdict would be too hasty. In fact, as we are going to see, the graph connecting these two variables has many more nodes to be taken into account.

In section \ref{frdr} we have seen that all cases of rotational frame-dragging follow the same scheme. The degree of Machianity of each case then followed from the particular physical setting considered. We start by giving information about the material distribution (the equation of state involving density $\rho$ and pressure $p$ of matter, which in turns determines the stress energy tensor $T_{\mu\nu}$). We then give the set of background symmetries (with related Killing vectorfields) compatible with $T_{\mu\nu}$ (for example, external cylindrical symmetry for the Einstein-Lense-Thirring case). We then compute the angular momentum via \eqref{lensth4a}, and from that we construct a linear perturbation $\delta h_{\mu\nu}$ of the background. We use the stress-energy tensor to calculate the perturbation via Einstein's field equations. In virtue of the symmetries of the background, we apply Noether's theorem to find that only the $0i$ $(i=1,2,3)$ part of the field equations are involved. We thus find the form of the dragging potential $f^a$ by \eqref{disp} (or the particular cases \eqref{lensth7} and \eqref{lensth9}) and, from this, we construct the vector potential $\mathbf{A}=(0,0,f^a)$ by fixing a convenient gauge (e.g. the Lorenz gauge). Finally, we get to the angular velocity of precession $\boldsymbol{\Omega}$ of a local inertial frame via \eqref{lensth5}, again by a choice of gauge that fixes the value of the function $e^\alpha$. Hence, the graph of the structural model is:
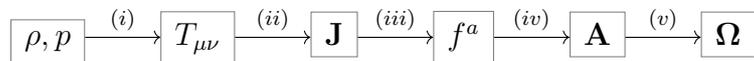
\begin{center}
\begin{tikzcd}[cells={nodes={draw=gray}}]
                                 \rho, p \arrow[r, black, "(i)"] & T_{\mu\nu} \arrow[r, black, "(ii)"] & \mathbf{J} \arrow[r, black, "(iii)"] & f^a \arrow[r, black, "(iv)"] & \mathbf{A} \arrow[r, black, "(v)"] & \boldsymbol{\Omega}\\
\end{tikzcd}
\end{center}
Of course, the actual structural equations relating these variables would be extremely complex. However, for our purposes it is sufficient to evaluate the mediating principles invoked at each step. They are:
\begin{itemize}
\item [(i)] State equation of matter (analytic functional dependence).
\item [(ii)] Equation \eqref{lensth4a} (analytic functional dependence) plus extra-theoretical (mathematical) symmetry considerations (e.g. Birkhoff theorem).
\item [(iii)] Einstein's field equations (law of nature), weak field approximation, Noether's theorem (mathematical law), equation \eqref{disp} plus boundary conditions (e.g. $\omega(\infty)=0$ or closeness).
\item [(iv)] Gauge fixing.
\item [(v)] Equation \eqref{lensth5} (analytic functional dependence), gauge fixing.
\end{itemize}

We submit that, given the treatment of rotational frame-dragging presented in section \ref{frdr}, the above model faithfully captures the relevant physical variables and their dependence for the phenomena in question. We also claim that the above model is much more powerful and useful for causal analysis purposes than standard counterfactual frameworks. In fact, our model can be used either to analyze the general type of dependence relations in the formal schema of rotational frame-dragging, which includes all cases from Einstein-Lense-Thirring effect to general FLRW Machian cosmologies, or to analyze the token relations in each of the particular cases just mentioned. For comparison, Lewis' theory can only handle token cases. The great enhancement in analytical power is evident when we try to assess the Machianity of the different occurrences of frame-dragging. First of all, the Machian hallmark of frame-dragging is given by the fact that the exogenous variables represent material degrees of freedom, making manifest Mach's idea of material origin of inertial effects. Moreover, all the considerations we have dispersed here and there in section \ref{frdr} are beautifully summarized in the list of mediating principles (i)-(v). The less Machian cases feature ``absolute'' boundary conditions in (iii), while more Machian cases feature a condition of closeness in (iii), or even no boundary condition at all.

Coming back to the causality question, it is now clear why the standard counterfactual analyses of the dependence relation between $\mathbf{J}$ and $\boldsymbol{\Omega}$ performed so far gave an indecisive verdict - and why we cannot equate frame-dragging to everyday instances of causation, such as the stone/window case. Basically, the standard approaches systematically overlook both the fine-grained linking structure between $\mathbf{J}$ and $\boldsymbol{\Omega}$ and the dependence chain upstream of $\mathbf{J}$. So what is the verdict delivered by the SEM framework? We note that all links except for (iii) are not strictly speaking justified by laws of nature. The state equation for matter might count as a law in so far as it constrains $\rho$ and $p$, but it just establishes a functional dependence between them and the stress-energy tensor. The abundance of non-lawlike mediating principles hints at the fact that the dependence involved is much more logical/metaphysical than nomological. However we cannot speak of a pure case of grounding here because the middle link (iii) involves Einstein's field equations. (The nomological link here ``corrupts'' the whole chain, in much the way that a single stochastic link in an otherwise deterministic mechanism makes the whole mechanism's operation stochastic.)  Furthermore, (iii) is the most important link in the chain, since it connects the upstream variables describing the overall material distribution with the downstream variables accounting for the (geometrical) properties of local inertial frames. Thus, we are in a situation where the ``heart'' of the chain depicts a nomological kind of dependence, while the rest of the links point at a metaphysical one. In our opinion, this mixed chain is the sign that a non-standard determination relation is at work here.

The other reason to view the determination relation here as not a purely causal relation has to do with the invertible nature of the connection between inertial (metrical) structure and matter distribution in GR. Although the specific structure we laid out above is not directly invertible (that is, one can't input an $\boldsymbol{\Omega}$ and derive the state of matter $\rho, p$ by working the mathematical derivations in reverse), it is nevertheless the case in GR that the metric field directly determines the material distribution via Einstein's equations. So one could presumably construct a closely related causal graph starting with a variable representation of the metric with more or less local precession and ending with a determinate quantity $\mathbf{J}$ of angular momentum in the appropriate coordinate gauge.  Once again, the field equations, which are lawlike, would play a key role in justifying one of the middle links of the graph, yet we would feel no temptation to consider the graph as establishing a causal connection between $\boldsymbol{\Omega}$ and $\mathbf{J}$.

The fact that the graph would link things that are simultaneous in the chosen coordinate frame or frames is not the reason why we would be reluctant; the explanation lies elsewhere, in deeply held physical intuitions about how spacetime structure and matter present in spacetime may relate to each other.  This is not the place to explore those intuitions, and whether they ought to be defended or questioned.  Our point is that the mathematical connections may be just as tight if we run them in the opposite direction, from a precessing metric to a quantity of angular momentum in the matter distribution about the central point.  Despite this, an intervention on the central metric entailing a concomitant change in distant angular momentum does not feel like a connection that reveals causation, but rather at most a lawlike co-dependence relation. If we judge by the similarity of the two dependence stories, we should judge them alike:  both as amounting to a non-standard determination relation that should not be held to be a clear case of causation.

We finish this section by considering a possible no-go objection against our reasoning. The objection goes like this: the frame-dragging effects we are discussing are just the result of an approximation, so we should not read too much into them, especially from a metaphysical perspective. To this objection we reply that the approximation works extremely well for physically realistic situations, where small velocities and weak gravitational fields are involved, to the point that no higher-order corrections are needed to achieve empirically adequate results (see \citealp{554}). This is, we think, more than a sufficient reason to investigate what kind of goings-on is captured by \eqref{disp}, especially in light of the fact that nobody, to our knowledge, has been so far able to come up with a higher-order derivation of frame-dragging effects in GR.

\section{What kind of dependence?}\label{five}
We have spent much of the previous section to establish what the dependence relation underpinning frame-dragging effects is not. It is not straightforwardly causal, but it has too much physical import to be considered just metaphysical. So it seems that we are looking at a strange new animal lurking behind the bushes. The question then is: what is a frame-dragging relation? As we are going to see, the answer to this question is not univocal and is open to debate. Roughly speaking, we can isolate two possible responses.

The first response, which we might call \emph{monist}, takes to its extreme consequences Wilson's unificatory thesis for dependence relations. Hence, this brand of monism would deny that the dependence depicted by frame-dragging is half-way between causation and grounding. This is because there is just \emph{one} dependence relation -- call it causation with a big ``c'', grounding with a big ``g'', or what have you -- so the metaphysics of windows shattering, that of singletons, and that of frame-dragging is one and the same. The reason why, so far, metaphysicians had the impression of dealing with conceptually different cases is that they focused too much on details that cut no ontological ice. From this perspective, the SEM framework is particularly useful in pointing out what -- according to monism -- went wrong in the standard analytical approach. In a nutshell, the issue boils down to the mediating principles justifying a structural model. To see this, let's go back to the stone/window and member/singleton cases. Both circumstances are modeled in the very same way: same variables, same structural equations, same counterfactual pattern instantiated. In other words, the SEM framework simply does not distinguish between the two dependencies. The only demarcation is given by the mediating principles involved, but such principles are \emph{external} to the analytical framework exploited. So, in some sense, mediating principles are just a ladder used to reach the structural model and then kicked off once the model is complete. As such, these principles have a pragmatic rationale but very dubious metaphysical import. Another cue that might lead in this direction is that there is no consensus over the metaphysical status of laws of nature, some philosophers arguing that they are just another brand of metaphysical principles (e.g. necessitarianism), some others going as further as claiming that there is no such thing as laws of nature. Clearly, monism is an appealing thesis in light of this debate, since it dispenses with the need of providing an account of laws of nature in order to spell out what a dependence relation is. On the other hand, monism has a huge downside in that it might be accused of solving the issue by trivializing it. Indeed, dismissing the prima facie huge conceptual differences among dependence relations as descriptive fluff seems a rather unsatisfactory move. Usually, we clarify things up by adding details to the description, not by blurring the whole picture.

Hence, one might consider an opposite, \emph{pluralist}, response to the original question. Pluralism differs from monism in that it takes the demarcation by mediating principles very seriously from a metaphysical point of view. Thus, under a pluralist reading of the SEM framework, causation and grounding are distinct but related concepts. This means that dependence relations are not sparse, but can be arranged in a scale depending on the type of principles mediating such dependencies. Hence, on one side sits causation, which is mediated by laws of nature only, while on the opposite side sits grounding, which is not mediated at all by laws of nature. In this picture, the claim that frame-dragging dependence lies between causation and grounding is literally true, since it is only partially mediated by laws of nature. In this sense, frame-dragging dependence might be (unimaginatively) dubbed ``caunding'' or something like that. The problem with this line of reasoning is that it does not clarify whether such a dependence scale has just three positions (nomological, mixed, metaphysical) or it covers a wider spectrum capturing the degree of mixing. In this second case, a further criterion should be provided for ``counting'' or ``weighing'' the lawlike links in a dependence chain, in order decide the position in the spectrum of the corresponding relation. Lacking this additional criterion, the question regarding the nature of frame-dragging dependence remains only partially answered. Contrary to monism, pluralism seeks indeed to provide a non-trivial, informative account of dependencies by adding more details to the picture. However, the added layer of complexity seems to bring new issues to the fore, rather than settling the already existing ones. From this point of view, it is contentious whether pluralism fares really better than monism.

\section{Conclusion}
Standard analyses of dependence relations are often aimed at well-behaved, clear-cut cases taken from everyday life. For example, all of us have a clear grasp of what it means for a lightning to determine a fire in the woods, or for a marble to be colored in virtue of being red. These analyses commonly exploit tools such as Lewisian counterfactuals, and rely on demarcation criteria given in terms of distinctions like events/facts, temporality/fundamentality, synchronicity/diachronicity, contingency/necessity. In this paper, we have tried to show that when it comes to fundamental physics, and GR in particular, things are not so straightforward. The lesson we have drawn from this fact is that, perhaps, we should follow a new path in metaphysical theorizing. The blazing of such a trail has recently started with the work of Schaffer, Wilson, and others on a generalized SEM framework that unifies the -- so far, disjoint --  analyses of causation and grounding. What we have done is basically to apply this framework to frame-dragging dependencies in GR, showing that it is possible to deliver a (partial) characterization of the underlying dependence relation, which is at least conceptually clearer than that delivered by standard analytical frameworks. Of course, much more has still to be said regarding such a relation in order to reach a full metaphysical characterization. However, as discussed in the previous section, this is part of the broader issue of making full metaphysical sense of the tools provided by the generalized SEM framework (in particular of the demarcation criterion in terms of the laws of nature/metaphysical principles distinction). As such, our work is just another preliminary step toward a full understanding of this non-standard framework. A further step would be to investigate the nature of the dependence relation underpinning geodesic motions in GR, looking for possible differences with the frame-dragging case. In the end, the hope is that a fully developed analysis of dependencies in spacetime physics might help in solving the thorny issue of providing a clear metaphysical story for the emergence of classical spacetime from an underlying quantum-gravitational regime. At this stage, however, fulfilling this latter task lies far in the future.



\pdfbookmark[1]{Acknowledgements}{acknowledgements}
\begin{center}
\textbf{Acknowledgements}
\end{center}
We are grateful to an anonymous referee and the editors of this volume for their comments on an earlier draft of this paper. Antonio Vassallo acknowledges financial support from the Spanish Ministry of Science, Innovation and Universities, fellowship IJCI-2015-23321.

\pdfbookmark[1]{References}{references}
\bibliography{biblio}
\end{document}